\documentclass[a4paper,12pt, leqno]{article}
\usepackage{natbib,enumerate,graphicx,array,mathtools}
\usepackage{amsmath}
\usepackage{amsthm}
\usepackage{amssymb}
\usepackage{hyperref}
\usepackage{amsfonts}
\usepackage{amscd}
\usepackage[utf8]{inputenc}
\usepackage{palatino}

\newtheorem{theorem}{Theorem}

\newtheorem{lemma}[theorem]{Lemma}

\newtheorem{proposition}[theorem]{Proposition}
\theoremstyle{definition}

\theoremstyle{remark}
\newtheorem{example}[theorem]{Example}

\newcommand* \ba {\mathop{\rm ba}\nolimits }

\newcolumntype{x}[1]{>{\centering\arraybackslash}p{#1}}
\newcolumntype{C}[1]{>{\centering\let\newline\\\arraybackslash\hspace{0pt}}m{#1}}
\usepackage{setspace}

\def\one{\boldsymbol{1}}
\newcommand{\acore}{\operatorname{acore}}

\def\R{\mathbb{R}}
\def\N{\mathbb{N}}
\def\A{\mathcal{A}}

\def\O{\Omega}
\def\o{\omega}

\def\:={\mathrel{\mathop:}=}

\newcommand* \core {\mathop{\rm core}\nolimits }

\title{On the Nonemptiness of the Core: \\Connecting Cooperative Games and Asset Pricing\thanks{Pintér acknowledges   support from the Hungarian Scientific Research Fund under   project K 146649 and by the HUN-REN under the project TKCS-2024/34.}}

\author{Bernard Cornet\thanks{Department of Economics, The University of Kansas, 1460 Jayhawk Boulevard, Lawrence 66045, KS, USA, cornet@ku.edu} \, and Mikl\'os Pint\'er\thanks{Corvinus University of Budapest and HUN-REN-BME-BCE Quantum Technology Research Group, pmiklos@protonmail.com.}}

\begin{document}

\maketitle
	
\begin{abstract}
\baselineskip=8pt
We show that the Bondareva--Shapley theorem for infinite cooperative games can be derived from a version of the fundamental theorem of asset pricing. The link is the Choquet extension of a game, viewed as a possibly nonlinear pricing rule on the space of measurable simple payoffs. Under this identification, balancedness translates into no arbitrage, and a linear functional dominated by the pricing rule yields an allocation in the anti-core; the core statement follows by duality. 

Applying the arbitrage-free pricing theorem of \citet{Cornet2025b}, we recover the characterization of the nonvacuity of the core initiated by \citet{Schmeidler1967} and established in full generality  by \citet{BartlPinter2023} for bounded-below games. Unlike the earlier derivation in \citet{Cornet2025b}, our argument does not require the game to have bounded variation or the collection of coalitions to be a $\sigma$-algebra.

\smallskip
		
\end{abstract}

\section{Introduction}

Two fundamental existence questions arise in cooperative game theory and in asset pricing. When does a cooperative game admit an allocation that no coalition can block? When does a possibly nonlinear pricing rule admit a compatible linear valuation? The answers are expressed in different languages, balancedness and nonemptiness of the core in the first case, no-arbitrage and linear support in the second, but they rest on the same underlying separation principle. This paper makes that connection explicit.

For finite cooperative games, the Bondareva--Shapley theorem \citep{Bondareva1963,Shapley1967} states that the core \citep{Shapley1955,Gillies1959} is nonempty if and only if the game is balanced. Its extension to infinite player sets has required a succession of increasingly general results; see \citet{Schmeidler1967,Kannai1969,Kannai1992,MarinacciMontrucchio2004,BartlPinter2023,BartlPinter2024}. In the formulation relevant here, a game defined on an arbitrary field of coalitions and bounded below has a nonempty core exactly when it is balanced \citep{BartlPinter2023}. 

In asset pricing, the corresponding principle is that the absence of arbitrage entails the existence of a linear valuation; see \citet{HarrisonKreps1979,DelbaenSchachermayer1994,DelbaenSchachermayer2006}. Building on \citet{ChateauneufCornet2022} and \citet{BastianelloChateauneufCornet2024}, \citet{Cornet2025b} establishes such a result for pricing rules that need not be linear or subadditive: subject to a local boundedness condition, the relevant no-arbitrage property is equivalent to the existence of a continuous linear functional dominated by the pricing rule.

Our observation is that the theorem characterizing the nonvacuity of the core follows directly from this asset-pricing result. Let $v$ be a game and let $\hat v$ denote its Choquet extension to the normed space $S(\Omega,\mathcal A)$ of measurable simple functions. Interpreting simple functions as contingent payoffs turns $\hat v$ into a pricing rule. Anti-balancedness of $v$ then implies the no-arbitrage condition for $\hat v$. The asset-pricing theorem provides a continuous linear functional below $\hat v$; through the duality between simple functions and bounded finitely additive set functions, this functional defines an element of the anti-core. Applying the argument to the dual game gives the usual core formulation.

This choice of payoff space is what permits the full generality of the argument. The derivation in Section~2.8 of \citet{Cornet2025b} uses bounded measurable payoffs and assumes both that the game has bounded variation and that $\mathcal A$ is a $\sigma$-algebra. Working with measurable simple payoffs instead, and applying Theorem~1 of \citet{Cornet2025b}, removes both assumptions: $\mathcal A$ may be an arbitrary field, and the game $v$ need only be bounded above. The counterexample in the final section confirms that this remaining condition cannot simply be omitted.

The rest of the paper is organized as follows. Section~2 states the two existence theorems and derives the cooperative-game result through the Choquet extension. Section~3 proves the auxiliary proposition used in that derivation. Section~4 concludes and examines the role of boundedness.

\section{From Cooperative Games to Asset Pricing}
\subsection{Nonvacuity of the core}

The aim of this paper is to show that  results in cooperative game theory can be derived from asset-pricing theory using the Choquet extension of a game. We focus on  the core \citep{Shapley1955,Gillies1959}, a central solution concept in cooperative games.


We first introduce some notation and concepts. Let
$\Omega$ be a nonempty set and let $\mathcal A$ be a field of subsets of
$\Omega$, so that $(\Omega,\mathcal A)$ is a chargeable space. In the language
of cooperative game theory, $\Omega$ is the set of players and $\mathcal A$ is
the collection of admissible coalitions. A transferable-utility cooperative game
(henceforth, simply a \emph{game}) is a set function
$v\colon\mathcal A\to\mathbb R$ satisfying $v(\emptyset)=0$, where $v(A)$ is
the value of coalition $A\in\mathcal A$. We denote by
$\mathcal G(\Omega,\mathcal A)$ the class of games and by
$\ba(\Omega,\mathcal A)$ the space of bounded additive games on
$(\Omega,\mathcal A)$.
For a game $v\in\mathcal G(\Omega,\mathcal A)$, its core and anti-core are,
respectively,
\begin{align*}
\core(v)
&\coloneqq
\bigl\{\mu\in\ba(\Omega,\mathcal A):
  \mu(A)\geq v(A)\text{ for all }A\in\mathcal A,
  \ \mu(\Omega)=v(\Omega)\bigr\},\\
\acore(v)
&\coloneqq
\bigl\{\mu\in\ba(\Omega,\mathcal A):
  \mu(A)\leq v(A)\text{ for all }A\in\mathcal A,
  \ \mu(\Omega)=v(\Omega)\bigr\}.
\end{align*}
For each $A\in\mathcal A$, we write $A^c\coloneqq\Omega\setminus A$. The dual
(or conjugate) game $v^*\colon\mathcal A\to\mathbb R$ is defined by
$
v^*(A)\coloneqq v(\Omega)-v(A^c)$ for 
$A\in\mathcal A.
$
Then the core and anti-core are related by

\smallskip \quad$
\acore(v^*)=\core(v)
\qquad\text{and}\qquad
\core(v^*)=\acore(v).
$
\smallskip

The Bondareva--Shapley theorem \citep{Bondareva1963,Shapley1967} gives
necessary and sufficient conditions for the nonemptiness of the core (and,
dually, of the anti-core). This finite-dimensional result has been extended to
several infinite-dimensional settings; see
\citet{Schmeidler1967,Kannai1969,Kannai1992,MarinacciMontrucchio2004,
BartlPinter2023,BartlPinter2024}. We follow the line of generalizations initiated
by \citet{Schmeidler1967} and developed in full generality by
\citet{BartlPinter2023}.

The following result states that a bounded-below game has a nonempty core if
and only if it is balanced. It also gives the corresponding dual characterization
of the nonvacuity of the anti-core. For $A\in\mathcal A$, we let $\one_A\colon\Omega\to\{0,1\}$ denote its indicator
function, with the convention $\one_\emptyset=0$. We also write $\mathbb N^*$
for the set of strictly positive integers. 

\begin{theorem}\label{thm:thm1}
Let $v\in\mathcal G(\Omega,\mathcal A)$ be a game.

\noindent  (a) If  the game $v$ is bounded below,   the following two conditions are equivalent:
     
$(i')$  $\core (v) \neq \emptyset$;
   
$(ii')$ $v$ is balanced, that is,

\quad$\forall K\in \N^* , \; \forall (\theta_1,\dots,\theta_K) \in \R^K_+, \; \forall (A_1,\dots,A_K)\in \A^K $

\quad $\sum_{k=1}^K \theta_k \one_{A_k}= \one_\Omega   \implies \sum_{k=1}^K \theta_k v({A_k}) \leq v(\Omega)  .
$

 \noindent (b) If the game $v$ is bounded above,   the following two conditions are equivalent:
     
$(i)$  $\acore (v) \neq \emptyset$;
  
$(ii)$ $v$ is anti-balanced, that is, 

\quad $\forall K\in \N^* , \; \forall (\theta_1,\dots,\theta_K) \in \R^K_+, \; \forall (A_1,\dots,A_K)\in (\A)^K$

\quad$\sum_{k=1}^K \theta_k \one_{A_k}= \one_\Omega   \implies \sum_{k=1}^K \theta_k v({A_k}) \geq v(\Omega).
$
\end{theorem}

Theorem \ref{thm:thm1} is proved in Section 2.3, as a consequence of the version of the 
so-called Fundamental Theorem of Asset Pricing (Theorem~\ref{thm:thm2}) that is presented in the next section.

\subsection{The Fundamental Theorem of Asset Pricing}
We next recall a version of the Fundamental Theorem of Asset Pricing (Theorem~1
of \citet[p.~716]{Cornet2025b}), which belongs to the tradition initiated by
\citet{HarrisonKreps1979}; see also
\citet{DelbaenSchachermayer1994,DelbaenSchachermayer2006}.

\begin{theorem}\label{thm:thm2}
Let $X$ be a Hausdorff locally convex topological vector space, let $X'$ be its
topological dual space,  let $f\colon X\to\mathbb R$ satisfy $f(0)=0$, and suppose that
$f$ is bounded above on a neighborhood of $0$. 

Then the following assertions
are equivalent:

$(i)$ $f$ satisfies the arbitrage-free condition \textup{(AF)}: 

\quad $\forall  K \in \N^*$, $\forall   (\theta_1,\dots,\theta_K) \in \R^{K}_+$, $\forall (x_1,\dots,x_K) \in X^K$ 

\quad$\sum_{k=1}^K \theta_k x_k = 0 \implies \sum_{k=1}^K\theta_k f (x_k) \geq 0 \, ;$

$(ii)$ there exists $\hat\mu\in X'$ such that $\hat\mu\leq f$; that is,
$\hat\mu(x)\leq f(x)$ for all $x\in X$.

\end{theorem}

Condition $(ii)$ is equivalent to the nonemptyness of   the subdifferential of $f$ at $0$, that is, $\partial f(0)\neq\emptyset$, since
$$
\partial f(0)
\coloneqq
\{x'\in X':x'(x)+f(0)\leq f(x)\, \forall x\in X\}
=\{x'\in X':x'\leq f\}.
$$
We briefly discuss the financial interpretation of this theorem. Hereafter, the set
 $X$ is either the normed space $S(\Omega,\mathcal A)$ of $\mathcal A$-measurable simple functions or 
$B(\Omega,\mathcal A)$, the Banach space of bounded
$\mathcal A$-measurable functions (when $\mathcal A$ is a $\sigma$-algebra).
The set $\Omega$ represents the states of nature at the uncertain date $t=1$
(tomorrow) in a two-date model, whereas date $t=0$ (today) is known and
$\mathcal A$ is the collection of admissible events.

A {\it payoff} is a bounded measurable function $x\colon\Omega\to\mathbb R$, with
$
\|x\|_\infty\coloneqq\sup_{\omega\in\Omega}|x(\omega)|<+\infty,
$
that specifies the flow of money at date $t=1$. If state $\omega$ occurs, then
$|x(\omega)|$ is paid by the agent when $x(\omega)<0$ and received by the agent
when $x(\omega)>0$.

The market assigns to each payoff $x\in X$ a date-$0$ price, or cost, $f(x)$ satisfying the sign convention that $|f(x)|$ is a gain when $f(x)<0$ and a payment when
$f(x)>0$. A functional $f\colon X\to\mathbb R$ is called a \emph{pricing rule}
provided only that $f(0)=0$, so that the null payoff has zero price.

Condition \textup{(AF)} says that any portfolio whose aggregate payoff tomorrow
is zero cannot have a gain today, that is, a negative aggregate cost. Interpreting $\hat\mu$ as
a stochastic discount factor, $\hat\mu(x)$ is the present value of the future
payoff $x$. The theorem therefore states that $f$ is arbitrage-free if and only
if there exists such a present-value functional $\hat \mu$, for which $\hat\mu(x)$ never
exceeds the market cost $f(x)$. For further discussion of condition
\textup{(AF)} and probabilistic interpretations of $\hat\mu$ and $f$, see
\citet{Cornet2025a,Cornet2025b}.

\subsection{From games to asset pricing  via the Choquet extension}

This section derives Theorem~\ref{thm:thm1}, the characterization
of the nonvacuity of the core, from Theorem~\ref{thm:thm2}, the version of
the Fundamental Theorem of Asset Pricing stated above.
 The bridge between the two settings is the Choquet
extension $\hat v:S(\Omega,\mathcal A)\to \mathbb R$ of a game $v\colon\mathcal A\to\mathbb R$, interpreted as a
pricing rule on the space $S(\Omega,\mathcal A)$ of measurable simple functions.

We recall the Choquet functional introduced by \citet{Choquet1953}. Endow
$S(\Omega,\mathcal A)$ with the supremum norm. For a game
$v\in\mathcal G(\Omega,\mathcal A)$, its Choquet extension
$\hat v\colon S(\Omega,\mathcal A)\to\mathbb R$ is defined as follows. If
$x\in S(\Omega,\mathcal A)$ has range
$\{x_1,\ldots,x_m\}$, where $x_1<\cdots<x_m$, then
$$
\hat v(x)
\coloneqq
x_1v(\Omega)
+\sum_{i=2}^m(x_i-x_{i-1})
v\bigl(\{\omega\in\Omega:x(\omega)\geq x_i\}\bigr).
$$
The following proposition allows  us to derive Theorem~\ref{thm:thm1} from
Theorem~\ref{thm:thm2}.

\begin{proposition}\label{lem:lem2}
Let $\hat v\colon S(\Omega,\mathcal A)\to\mathbb R$ be the Choquet extension of $v\in\mathcal G(\Omega,\mathcal A)$.
\begin{enumerate}[(i)]
\item If  the game $v \colon \A \to \R$ is bounded above by $M\ge 0$, then 

\smallskip\quad $\hat v(x) \le || x ||_{\infty} (| v(\O) | + 2 M)$ for all $x\in S(\O,\A)$,

\smallskip  hence
$\hat{v}$ is bounded above on every bounded set, and therefore on some bounded neighborhood of $0$;

\item $\partial\hat v(0)\subseteq\acore(v)$;
\item If $v$ is anti-balanced, then $\hat v$ satisfies \textup{(AF)}.
\end{enumerate}
\end{proposition}

The pair $(S(\Omega,\mathcal A),\ba(\Omega,\mathcal A))$ is a dual pair; see,
for example, \citet[Section~5.14]{AliprantisBorder2006}. We therefore identify a bounded
charge $\mu\in\ba(\Omega,\mathcal A)$ with the corresponding continuous linear
functional on $S(\Omega,\mathcal A)$, given by
$
x\longmapsto\int_\Omega x\,\mathrm d\mu.
$
Under this identification, it is meaningful to compare
$\partial\hat v(0)$ and $\acore(v)$.


\begin{proof}[Proof of Theorem~\ref{thm:thm1}] $\bullet$ {\it Part~$(b)$.} We
apply Theorem~\ref{thm:thm2} to $X=S(\Omega,\mathcal A)$, a normed space under the sup norm, hence Hausdorff  locally convex, \hbox{and
$f=\hat v$.}

 \emph{(i) $\Rightarrow$ (ii).}\footnote{Note that this implication does not require $v$ to be bounded
above.}
Let $\mu\in\acore(v)$. If
$\sum_{k=1}^K\theta_k\one_{A_k}=\one_\Omega$ then additivity of $\mu$ and the
inequalities $\mu(A_k)\leq v(A_k)$ yield $v$  anti-balanced since

\smallskip\quad $
v(\Omega)=\mu(\Omega)
=\sum_{k=1}^K\theta_k\mu(A_k)
\leq\sum_{k=1}^K\theta_kv(A_k).$

\smallskip \emph{(ii) $\Rightarrow$ (i).}
Suppose that $v$ is anti-balanced and bounded above. By Parts~\emph{(i)} and \emph{(iii)} of Proposition~\ref{lem:lem2}, the hypotheses of
Theorem~\ref{thm:thm2} are satisfied. Hence, 
$\emptyset\neq \partial\hat v(0)\subseteq \acore(v)$ by Part~\emph{(ii)} of
Proposition~\ref{lem:lem2}.

\smallskip\noindent$\bullet$  \emph{Part (a)}. Apply Part~\emph{(b)}  to the dual game $v^*$.
Indeed, $v$ is bounded below if and only if $v^*$ is bounded above; moreover,
$v$ is balanced if and only if $v^*$ is anti-balanced, and
$\core(v)=\acore(v^*)$.
\end{proof}

\section{Proof of Proposition \ref{lem:lem2}}
We prepare  the proof of Proposition \ref{lem:lem2} with the following lemma.

\begin{lemma}\label{lem:lem1}
Let $v \in \mathcal{G}(\O,\A)$. Then the following two assertions are equivalent:

1. $v$ is anti-balanced;

2. $\forall  K\in \N^*$, $\forall   (\theta_1,\dots,\theta_K, \theta_{K+1}) \in \R^{K}_+ \times \R$, $\forall (A_1,\dots,A_K)\in \A^K$ 

\smallskip\quad
$\sum_{k=1}^K \theta_k \one_{A_k} + \theta_{K+1}  \one_\Omega = 0 \implies \sum_{k=1}^K\theta_k v({A_k})+ \theta_{K+1} v(\Omega)\ge 0 \, .$
\end{lemma}

\begin{proof}{ }
[1. $\implies$ 2.]\, Let $K \in \N^*$, $(\theta_1,\dots,\theta_K, \theta_{K+1}) \in \R^{K}_+\times \R$, and $(A_1,\dots,A_K)\in \A^K$ satisfy $\sum_{k=1}^K \theta_k \one_{A_k}+ \theta_{K+1}\one_\Omega = 0$. We distinguish two cases.

\smallskip\noindent
$\bullet$ Suppose that $\theta_{K+1} \geq -1$. Then $\sum_{k=1}^K \theta_k \one_{A_k} + (\theta_{K+1} + 1)\one_\Omega = \one_\O$ and $\theta_{K+1} + 1 \geq 0$. Since $v$ is anti-balanced,
$\sum \limits_{k=1}^{K} \theta_k v (A_k) + (\theta_{K+1} + 1)  v (\Omega) \geq v (\O) \, .
$
Therefore, $\sum_{k=1}^{K} \theta_k v (A_k) + \theta_{K+1}  v (\Omega) \geq 0$.

\smallskip \noindent
$\bullet$ Suppose now 
$\theta_{K+1} < -1$. Then $\sum_{k=1}^K -(\theta_k/\theta_{K+1}) \one_{A_k}=\one_\Omega$ and $-\theta_k / \theta_{K+1} \geq 0$ for every $k=1,\ldots,K$. Since $v$ is anti-balanced, $\sum_{k=1}^K -(\theta_k / \theta_{K+1}) v(A_k) \geq v(\Omega)$.
Thus $\sum_{k=1}^K  \theta_k  v({A_k}) + \theta_{K+1} v(\Omega) \geq 0$.

\smallskip\noindent
[2. $\implies$ 1.]\, Take $K \in \N^*$, $(\theta_1,\dots,\theta_K) \in \R^{K}_+$, and $(A_1,\dots,A_K)\in \A^K$ such that $\sum_{k=1}^K \theta_k \one_{A_k} = \one_\Omega$, and let $\theta_{K+1}:=-1$. Then $\sum_{k=1}^K \theta_k \one_{A_k} - \one_\Omega = 0$. Assertion~2 gives $\sum_{k=1}^K \theta_k v(A_k) - v(\Omega) \geq 0$, thus $\sum_{k=1}^K \theta_k v(A_k) \geq v(\Omega)$.
\end{proof}

\begin{proof}[Proof of Proposition \ref{lem:lem2}] 
$Part \, (i)$. Let $M \ge 0$ be an upper bound for $v$.
Let $x \in S(\Omega, \A)$ have range  $\{x_1,\ldots,x_K\}$ where $x_1 <  \ldots < x_K$. Then
\begin{align*}
\hat v(x) &\leq | x_1 v(\Omega) | + \sum \limits^{K}_{k=2}  (x_{k} - x_{k-1}) v(\{\o \colon x (\omega) \geq x_k \})  \\
&\leq  | x_1 v (\O) |  + \sum \limits^{K}_{k=2}  (x_{k} - x_{k-1}) M\\  &\leq || x ||_{\infty} | v (\O) | + 2 || x ||_{\infty} M 
= || x ||_{\infty} (| v(\O) | + 2 M).
\end{align*}
 $Part \,(ii)$. Let $\hat{\mu} \in X' :=S(\Omega, \A)'$  satisfy $\hat{\mu} \leq \hat{v}$. The map from $\ba(\Omega,\A)$ onto $S(\Omega,\A)'$ given by
$\mu\mapsto\hat\mu$, where $\hat\mu(x)=\int_\Omega x\,\mathrm d\mu$, is an isometry. Thus
$\mu(A)=\hat\mu(\one_A)\leq\hat v(\one_A)=v(A)$ for every $A\in\A$. Moreover,\break
$\hat\mu(\one_\Omega)\leq\hat v(\one_\Omega)$ and
$-\hat\mu(\one_\Omega)=\hat\mu(-\one_\Omega)\leq\hat v(-\one_\Omega)=-\hat v(\one_\Omega)$.
Consequently, $\mu(\Omega)=\hat\mu(\one_\Omega)=\hat v(\one_\Omega)=v(\Omega)$, and hence $\mu\in\acore(v)$.

\smallskip\noindent
 $Part \, (iii)$. Let $K\in \N^*$, $(\theta_1, \dots , \theta_K)\in \R_+^K$, and  $(x_1, \dots ,x_K) \in S(\Omega, \A)^K$  satisfy $\sum_{k=1}^K \theta_k x_k = 0$. Write the range of each $x_k$ as $\{x_k^1,\ldots,x_k^{m_k}\}$ where $x_k^1 < \ldots < x_k^{m_k}$. Then $x_k = x_k^1 \one_{\Omega}  +  \sum \limits^{m_k}_{i=2}  (x_k^{i} - x_k^{i-1}) \one_{\{\omega \colon x_k (\omega) \geq x_k^i \}}$, $k=1,\ldots,K$.
 
Thus, $ \sum^{K}_{k=1} \theta_k x_k = 0$ implies that 
\begin{equation*}
\sum^{K}_{k=1} \theta_k x_k^1 \one_\Omega + \sum^{K}_{k=1} \sum^{m_k}_{i = 2}  \theta_k \big(x_k^{i} - x_k^{i - 1 }\big) \one_{\{\omega \colon x_k (\omega) \geq x_k^i \}} = 0 \, .
\end{equation*}
By Lemma \ref{lem:lem1} and the anti-balancedness of $v$:
  \begin{equation*}
\sum^{K}_{k=1} \theta_k \hat{v} (x_k) = \sum^{K}_{k=1} \theta_k \left( x_k^1 v(\Omega)  + \sum^{m_k}_{i=2} \big(x_k^i - x_k^{i - 1}\big) v\big(\{\omega \colon x_k (\omega) \geq x_k^i \} \big)\right) \geq 0 \, .
\end{equation*}
Therefore, $\hat{v}$ satisfies the arbitrage-free condition  AF. 
\end{proof}

\section{Conclusion}

We have established a direct bridge between cooperative game theory and mathematical finance. More precisely, the characterization of arbitrage-free pricing rules in \citet{Cornet2025b} implies the infinite-dimensional version of the Bondareva--Shapley theorem. Choosing for payoff space $X:=S(\O,\A)$ yields the full generality obtained by \citet{BartlPinter2023}. By contrast, the argument in Section~2.8 of \citet{Cornet2025b}, taking for payoff space $X:=B(\O,\A)$,  requires stronger assumptions: $v$ needs to have bounded variation and $\A$ to be a $\sigma$-algebra. 

We hope that this connection will lead to further interactions between the two fields. In particular, it may provide new insights into both cooperative game theory and mathematical finance and facilitate further transfers of ideas and techniques between them.

Finally, \citet{BartlPinter2023} provide an example of a balanced game that is bounded above but not bounded below and whose core is empty (Example 5, pp. 156–157, in \citet{BartlPinter2023}). Therefore, one-sided boundedness cannot be dropped in \citet{Cornet2025b} either. The following example is simple and interesting because it demonstrates that even a game that is unbounded above can be balanced:

\begin{example}
Let \(\Omega=\mathbb{N}\), and let $\mathcal{A} = \bigl\{A\subseteq \Omega \colon A \text{ is finite or } A^c \text{ is finite}\bigr\}$ be the finite--cofinite algebra. Define $v\colon\mathcal{A} \to \mathbb{R}$ by
$v(A)= |A|,$ if $A$ is finite and  $v(A)= -|A^c|,$ if $A^c$ is finite. In particular, \(v(\emptyset)=v(\Omega)=0\). 

It is easy too see that $v$ is additive, but not bounded (neither above nor below), that is, $v \notin \ba (\Omega,\mathcal{A})$. Moreover, it also clear that $v$ is balanced (every additive set function is balanced), but its core is empty.
\end{example}





\end{document}